\documentclass[a4,showpacs,twocolumn,floatfix,aps
]{revtex4}
\usepackage{epsfig}
\usepackage{subfigure}

\newcommand{\beq}{\begin{equation}}  
\newcommand{\eeq}{\end{equation}}  
\newcommand{\bea}{\begin{eqnarray}}  
\newcommand{\eea}{\end{eqnarray}}

\begin{document}

\title{Long wavelength instability of coherent structures in plane Couette flow}
\author{Konstantin Melnikov$^1$}

\author{Tobias Kreilos$^{1,2}$}

\author{Bruno Eckhardt$^{1,3}$}
\affiliation{$^{1}$Fachbereich Physik, Philipps-Universit\"at Marburg, 
D-35032 Marburg, Germany}
\affiliation{$^{2}$Max Planck Institut f\"ur Dynamik und Selbstorganisation,
D-37077 G\"ottingen, Germany}
\affiliation{$^{3}$J.M. Burgerscentrum, Delft University of Technology, Mekelweg 2, 2628 CD Delft, The Netherlands}

\date{\today}

\begin{abstract}
We study the stability of coherent structures in plane Couette flow against
long-wavelength perturbations in wide domains that cover several 
pairs of coherent structures. For one and two pairs of vortices,
the states retain the stability properties of the small domains, 
but for three pairs new unstable modes are found. 
They are shown to be connected to bifurcations that break the
translational symmetry and drive the 
coherent structures from the spanwise extended state to a modulated one that is
a precursor to spanwise localized states. Tracking the stability of the orbits as
functions of the spanwise wave length reveals a rich variety of additional
bifurcations.
\end{abstract}


\maketitle

\section{Introduction}

Several shear flows including pipe flow and plane Couette flow show a transition to
turbulence while the laminar profile is still stable \cite{Grossmann2000}.
The transition has been connected
to the appearance of three-dimensional coherent structures that form in
subcritical bifurcations in the state space of the system 
\cite{Eckhardt2007, Eckhardt2008b, Kerswell2005, Nagata1990, Clever1997, Faisst2003, Wedin2004}.
The subsequent evolution of these bifurcations then shows secondary bifurcations
and boundary crisis, in which attractors break up and chaotic saddles form
\cite{Kreilos2012, Mellibovsky2011,Mellibovsky2012}.
In small domains, these phase space structures have been characterized
in some detail and close connections to the properties of
chaotic saddles in low-dimensional chaotic systems
have been found
\cite{Nagata1990, Clever1997, Faisst2003, Eckhardt2008a, Wedin2004, Schneider2008}.
A chaotic trajectory meanders between these states and transient approaches
to invariant states have been observed in numerics and experiments \cite{Hof2004,Tutty2007,Schneider2007a}.
The states from the small domain can be carried over to the wider domain by
periodic continuation. However, the additional spatial degrees of freedom then
allow for further spatial patterning and the development of localized structures,
\cite{Schneider2010a,Marinc2010,Mellibovsky2009,Avila2013,Duguet2009,Duguet2012}. 
The elementary building blocks of both localized and extended coherent
states are downstream vortices and streaks. Localized states come with various numbers
of vortices and streaks, a phenomenon that can be rationalized within the
snaking scenario in subcritical bifurcations
\cite{Schneider2010, Knobloch2008}.

Our aim here is to explore the connection between extended and
localized structures through the analysis of long-wavelength
instabilities of a particular class of states in plane Couette flow.
We focus on edge states, the relative attractors in the edge that
separates laminar and turbulent dynamics, because
they are relatively easy to find and to analyze, thanks to their
distinguished dynamical properties \cite{Toh2003,Skufca2006,Schneider2007b,Schneider2008}. 
Searches for other states usually require more
sophisticated approaches, such as the embedding into a family of 
flows that show linear instabilities
\cite{Nagata1990,Waleffe1998,Waleffe1997,Faisst2003,Wedin2004,Faisst2000}
or the tracking of states through secondary bifurcations
\cite{Pringle2007,Pringle2009}. The advantage of
edge tracking is that the states can be found by a direct time-integration
and that their location on the boundary between laminar and turbulent
dynamics gives them an immediate dynamical relevance. The states we study
here are edge states which we can trace from small to wider domains.
We study their stability against infinitesimal long-wavelength perturbations
and the properties of the new states that appear in bifurcations.

The outline of the paper is as follows. In section \ref{sec:GeneralSetting} the general setting of this work is presented. 
In section \ref{sec:Stability} we discuss the linear stability of spatially extended coherent states. In section \ref{sec:Domain3pi} an extended coherent state
that shows a long wavelength instability is discussed in detail. We show that the instability is connected to a symmetry breaking bifurcation
that drives the coherent structures from a spanwise extended state to a modulated one. In section \ref{sec:WideBoxes} we consider
even wider boxes and show that this modulated state evolves towards a spatially localized one. 
We conclude with a summary and an outlook in section \ref{sec:Conclusion}.

\section{General setting}
\label{sec:GeneralSetting}
We study coherent structures in incompressible plane Couette flow. The two infinite 
plates are parallel to the $x$-$z$ plane and move
in opposite directions along the $x$ axis with the velocity $\pm u_{0}$. We refer to $x$, $y$, and $z$ as streamwise, 
wall-normal and spanwise directions, respectively. The plates are located at $y=\pm d$.
The motion of the fluid can be described by the incompressible Navier-Stokes and continuity equations,
\begin{equation}
 \label{equation:NavierStokesDimensionless}
  \partial_{t} \textbf{u} + ( \textbf{u} \cdot \nabla ) \textbf{u}  = - \nabla p + \frac{1}{Re} \Delta \textbf{u}, \qquad \nabla \cdot \textbf{u} = 0
\end{equation}
for the velocity field $\textbf{u}(\textbf{x},t)=[u,v,w](x,y,z,t)$. The Reynolds number is defined as $Re=u_{0} d / \nu$ where
$d$ is half the distance between the plates and $\nu$ is the kinematic viscosity, so that in these dimensionless units 
the boundary conditions in the velocity field are $u_{x}(y=\pm 1)=\pm 1$ and all other components vanish.

Periodic boundary
conditions are applied in streamwise and spanwise directions and no-slip boundary conditions at the walls.
For the direct numerical simulations we use the program package \textit{channelflow} \cite{cfmanual} developed by 
J.F. Gibson. The integrator is based on a spectral discretization in the spatial directions, Fourier
modes are used in spanwise and streamwise directions and Chebyshev modes in wallnormal direction. For 
time stepping a 3rd-order semi-implicit backwards differentiation algorithm is applied. Fixed points,
periodic orbits and traveling waves are calculated with a Newton-Krylov hookstep
algorithm \cite{Viswanath2007}. The linear stability of the solutions is characterized
by eigenvalues and eigenfunctions which are computed with Arnoldi iteration.
Parameter continuation of solutions is computed with a predictor-corrector method using
a two step process: solutions are first quadratically extrapolated in pseudo-arclength, and 
then refined with a Newton-Krylov hookstep algorithm in a second step.
%
The numerical resolution used is $48 \times 33 \times 48$ modes for a computational domain with spatial 
dimensions $L_{x}\times L_{y} \times L_{z} = 2\pi \times 2 \times \pi$, and suitable refined for larger
domains. This resolution was found to be high enough to resolve all relevant structures
in the wall-normal direction that are important for the coherent structures and the strong velocity gradients. Almost
all calculations are performed for $Re=400$, the only exception being the continuation in Reynolds number in Fig 3 below.

Most flow fields discussed in this work show some kind of discrete symmetry \cite{Schmiegel1999,Gibson2008a}. 
Using the notation of \cite{Gibson2008a} we introduce symmetry operators of the general form
\bea
  \sigma [u,v,w](x,y,z) = \nonumber \\ (s_{x} u, s_{y} v, s_{z} w) (s_{x} x + a_{x} L_{x},
  s_{y} y, s_{z} z + a_{z} L_{z})
\eea
where the coefficients $s_{x}$, $s_{y}$ and $s_{z}$ can be either $1$ or $ -1$, leading to reflections and rotations.
The variables $a_{x}$ and $a_{z}$ are translations in units of the box dimensions $L_{x}$ and $L_{z}$ 
along the $x$ and $z$ axis, respectively.



\begin{figure}

\includegraphics[width=0.5\textwidth]{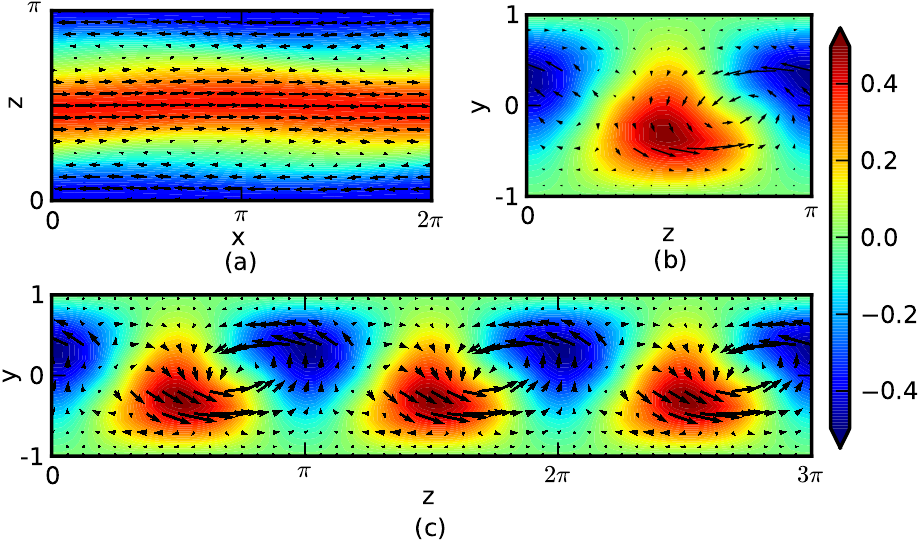}

\caption[]{(Color online) Different representations of the coherent state at $Re=400$. The panels show sections in 
(a) downstream and spanwise direction at midplane and in (b) spanwise and normal direction at $x=0$. Panel (c) shows the
embedding of the state (b) from a small domain in one three times as wide. The in-plane components
of the velocity are indicated by arrows, the perpendicular ones are color coded. }
\label{3xBoxYZ}
\end{figure}


\begin{table}
\begin{center}
    \begin{ruledtabular}
    \begin{tabular}{ll}
    n & $\mu$  \\ \hline 
    \multicolumn{2}{c}{$L_{x}$ $\times$ $L_{y}$ $\times$ $L_{z}$ = $2\pi \times 2 \times \pi$} \\
    \hline
    
    1 & 0.047394 \\
    2 & -1.20 $\cdot$ $10^{-7}$ \\
    3 & -6.63 $\cdot$ $10^{-7}$ \\
    \hline
     \multicolumn{2}{c}{$L_{x}$ $\times$ $L_{y}$ $\times$ $L_{z}$ = $2\pi \times 2 \times 2\pi$} \\
    \hline
     1 & 0.047394  \\
     2 & 3.31 $\cdot$ $10^{-6}$ \\
     3 & 1.10 $\cdot$ $10^{-7}$ \\
         \hline
     \multicolumn{2}{c}{$L_{x}$ $\times$ $L_{y}$ $\times$ $L_{z}$ = $2\pi \times 2 \times 3\pi$} \\
    \hline
     1 & 0.047394 \\ 
     2,3 & 0.004331 \\ 
     4,5 & 5.90 $\cdot$ $10^{-8}$ \\ 
    \hline
     \multicolumn{2}{c}{$L_{x}$ $\times$ $L_{y}$ $\times$ $L_{z}$ = $2\pi \times 2 \times 4\pi$} \\
    \hline
     1 & 0.047394    \\ 
     2,3 & 0.019090 \\ 
     4 & 2.05 $\cdot$ $10^{-7}$   \\ 
     5 & -6.90 $\cdot$ $10^{-7}$  \\
    \hline
     \multicolumn{2}{c}{$L_{x}$ $\times$ $L_{y}$ $\times$ $L_{z}$ = $2\pi \times 2 \times 5\pi$} \\
    \hline
    1 & 0.047394  \\ 
    2,3 & 0.027441 \\ 
    4 & 4.45 $\cdot$ $10^{-7}$  \\ 
    5 & 5.45 $\cdot$ $10^{-8}$  \\ 
    \hline
     \multicolumn{2}{c}{$L_{x}$ $\times$ $L_{y}$ $\times$ $L_{z}$ = $2\pi \times 2 \times 6\pi$} \\
    \hline
    1 & 0.047394  \\ 
    2,3 & 0.032704 \\ 
    4,5 & 0.004331 \\ 
    6 & 5.54 $\cdot$ $10^{-7}$  \\
    7 & 1.18 $\cdot$ $10^{-7}$ \\
    \hline
     \multicolumn{2}{c}{$L_{x}$ $\times$ $L_{y}$ $\times$ $L_{z}$ = $2\pi \times 2 \times 7\pi$} \\
    \hline
    1 & 0.047394  \\ 
    2,3 & 0.036193 \\ 
    4,5 & 0.012934  \\  
    6 & 1.04 $\cdot$ $10^{-7}$ \\
    7 & -6.44 $\cdot$ $10^{-7}$  
    \end{tabular}{}
    \end{ruledtabular}
\end{center}
\caption{\label{1xBoxStabilityTable} The leading eigenvalues $\mu$ for boxes with an increasing number of fundamental
vortices. Numerically, the eigenvalues of the two neutral eigenvalues that correspond to translations 
in $x$ and $z$ directions are  $\mathcal{O}{(10^{-7})}$, which provides a measure of the uncertainty
in the eigenvalue calculations.  The positive eigenvalues are well above zero and thus reliable. 
Some eigenvalues came in pairs of complex conjugate
ones with imaginary parts that are $\mathcal{O}{(10^{-8})}$ and smaller, so that within numerical accuracy
all eigenvalues are real.}
\end{table}

\section{Stability of steady states}
\label{sec:Stability}
In order to see how the linear stability properties change if the domain width is increased we calculate eigenvalues
for several characteristic steady states. The steady states are multiple copies of the state first
described by Nagata \cite{Nagata1990} and Clever and Busse \cite{Clever1997} and later characterized in further detail
by Waleffe \cite{Waleffe1998,Waleffe1997}. 
If this state is continued to lower Reynolds numbers then an upper branch can be detected. The upper and the
lower branches are born in a saddle-node bifurcation which in our geometry occurs at $Re=163.8$ \cite{Kreilos2012}.
Both branches are fixed points. The upper branch is linearly stable close to the saddle-node bifurcation while the
lower branch has one unstable eigenvalue. We start our analysis with the state from the lower branch
at $Re=400$ where it still has only one unstable direction, and where it is an edge state.
The flow field of this state is dominated by one vortex pair, see the contour plots in Fig.\,÷÷\ref{3xBoxYZ}.
For the wide domains we study periodically continued versions of this state by placing copies of the state
side by side in $z$-direction.

The eigenvalues for an increasing number of vortex pairs are given in Table \ref{1xBoxStabilityTable}, where we list
all positive eigenvalues and the two neutral ones. The neutral eigenvalues correspond to translations 
along the invariant $x$- and $z$-directions of plane Couette flow.
Their numerical value, which analytically should be zero and here is $\mathcal{O}{(10^{-7})}$,
indicates the numerical accuracy and shows that the positive eigenvalues are well above zero and numerically reliable.
Note also that domains that are multiples of smaller domains inherit some of the eigenvalues to within 
numerical accuracy. For instance, the leading eigenvalue $0.047394$ is shared by all domains,
the corresponding eigenfunction being perpendicular to the edge of chaos, pointing towards the laminar or turbulent attractor.
The
next to leading eigenvalue $0.004331$ for $L_z=3\pi$ reappears for $L_z=6\pi$. Of particular relevance here
is the appearance of more positive eigenvalues for larger domains.
For domains with one and two pairs of vortices the number of unstable eigenvalues is one, the stable manifold has co-dimension
one, and the states remain edge states, as in the small domains.  For three pairs of vortices a pair of degenerate
eigenvalues appears, and for more than three pairs of vortices even more unstable eigenvalues are
detected, as listed in Table \ref{1xBoxStabilityTable}. 



\begin{figure}
\includegraphics[width=0.5\textwidth]{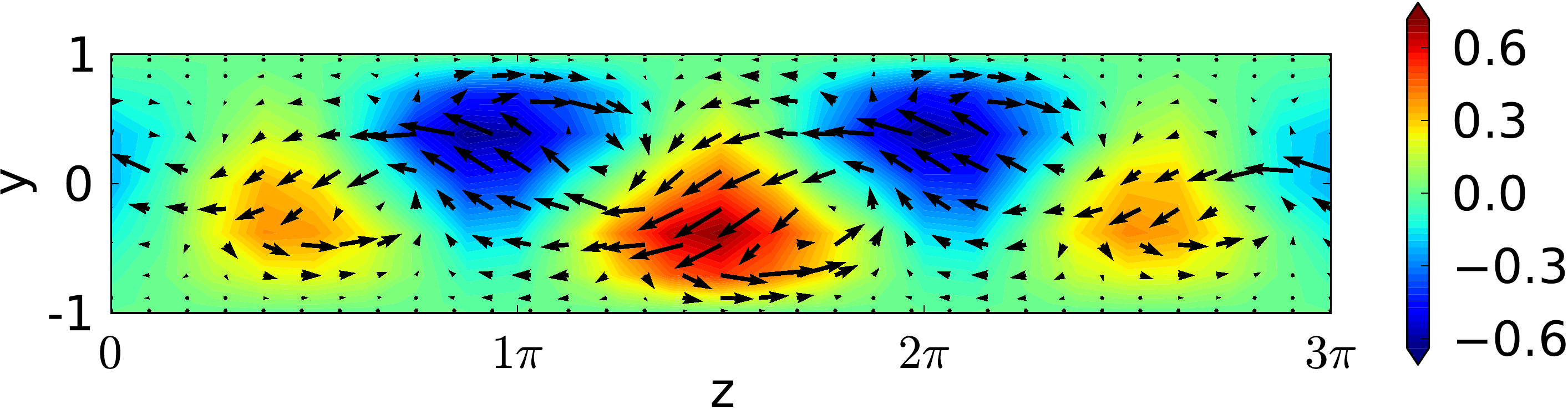}
\caption[]{(Color online) Contour plot of the traveling wave that emerges by symmetry breaking
from the stationary state in a domain with $L_z=3\pi$ and takes over as an edge state.  
Shown are sections perpendicular to the flow direction at $x=0$, with the
in-plane velocity indicated by vectors and the perpendicular component 
indicated in color.}
\label{TWPlot}
\end{figure}

\begin{figure}
\centering
\subfigure[]{\includegraphics[width=0.47\textwidth]{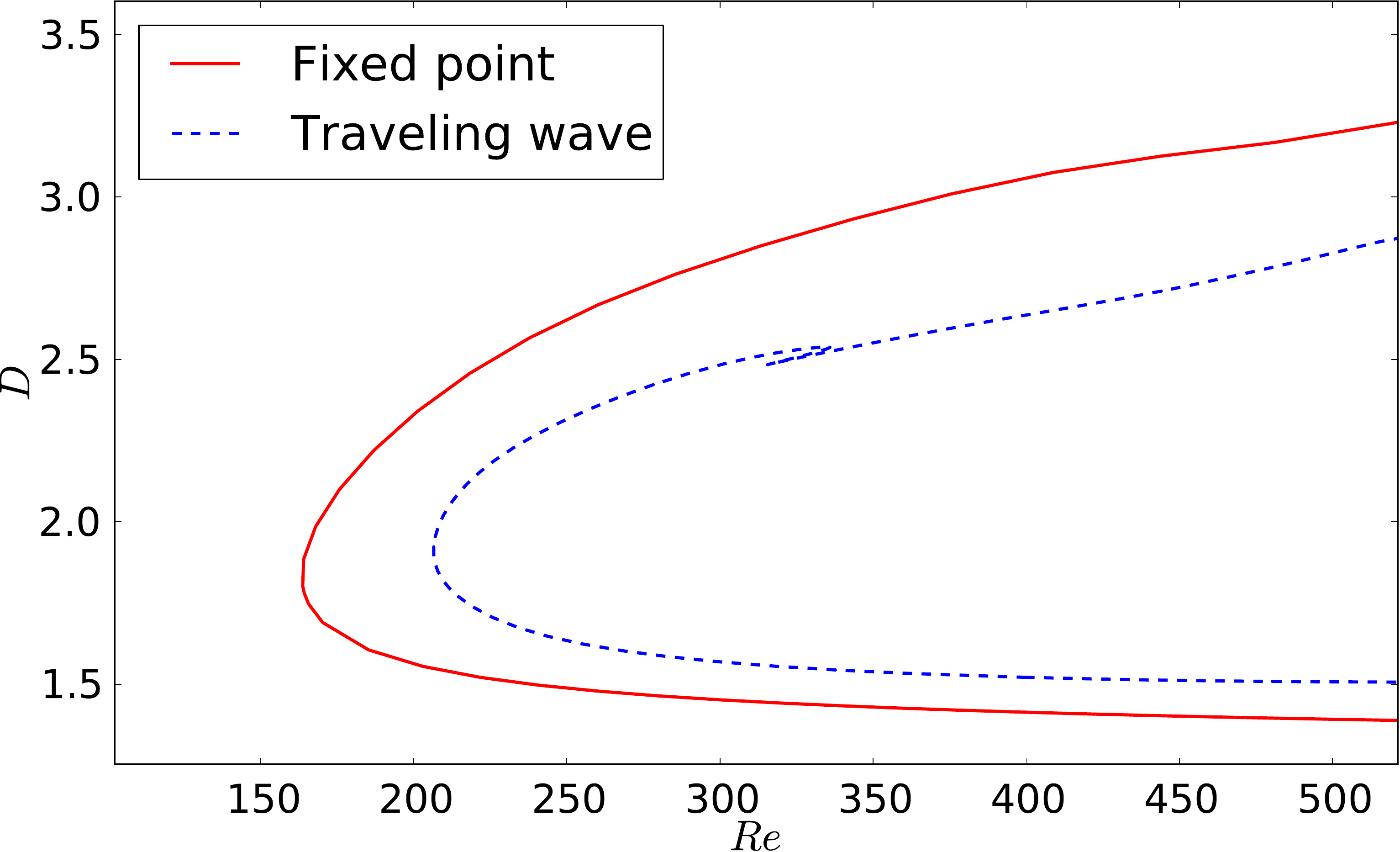}\label{ContReFP_TW}}
\subfigure[]{\includegraphics[width=0.23\textwidth]{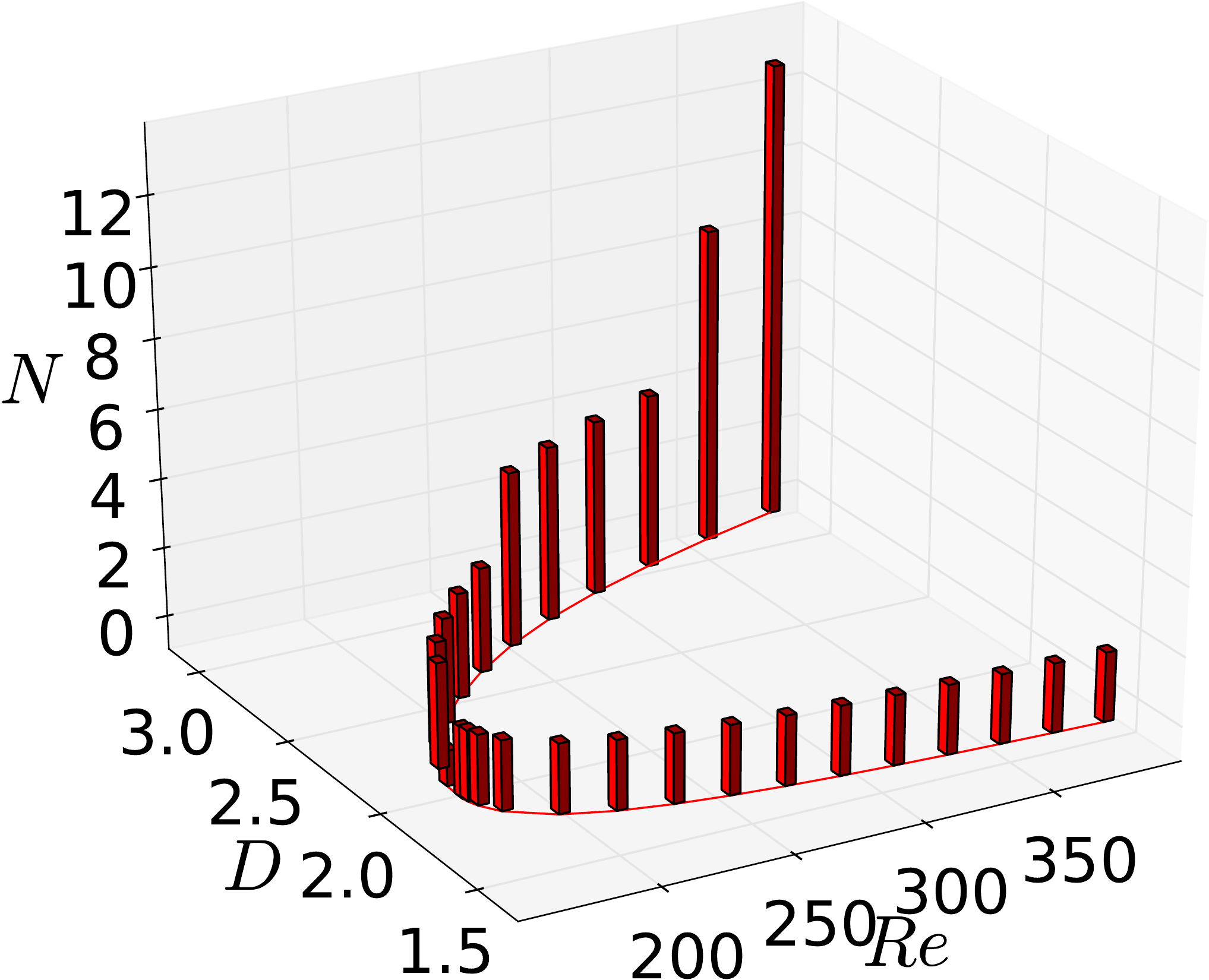}\label{FPStab3D}}
\subfigure[]{\includegraphics[width=0.23\textwidth]{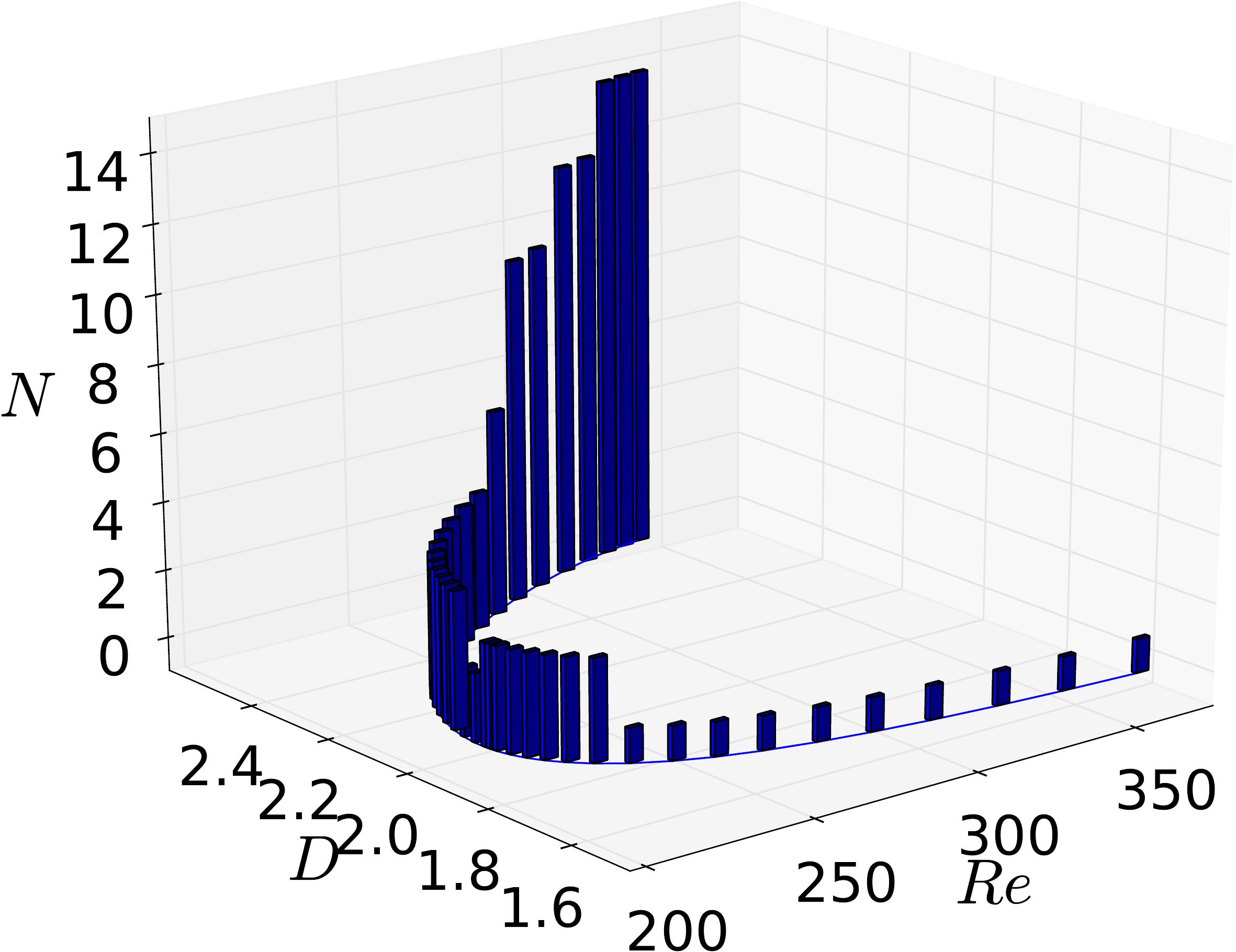}\label{TWStab3D}}
\caption[]{(Color online) Exact coherent states (a) as a function of Reynolds number 
and the number of unstable directions for (b) the stationary state and (c) the traveling wave in
a domain with $L_z=3\pi$. The histograms with the number of unstable
directions show that the lower branch states typically have one or two unstable
directions for most Reynolds numbers whereas the upper branch solution 
undergoes further bifurcations that bring in more unstable directions. }
\label{fig:ContinuationInRe}
\end{figure}

\begin{figure}
\subfigure[]{\includegraphics[width=0.47\textwidth]{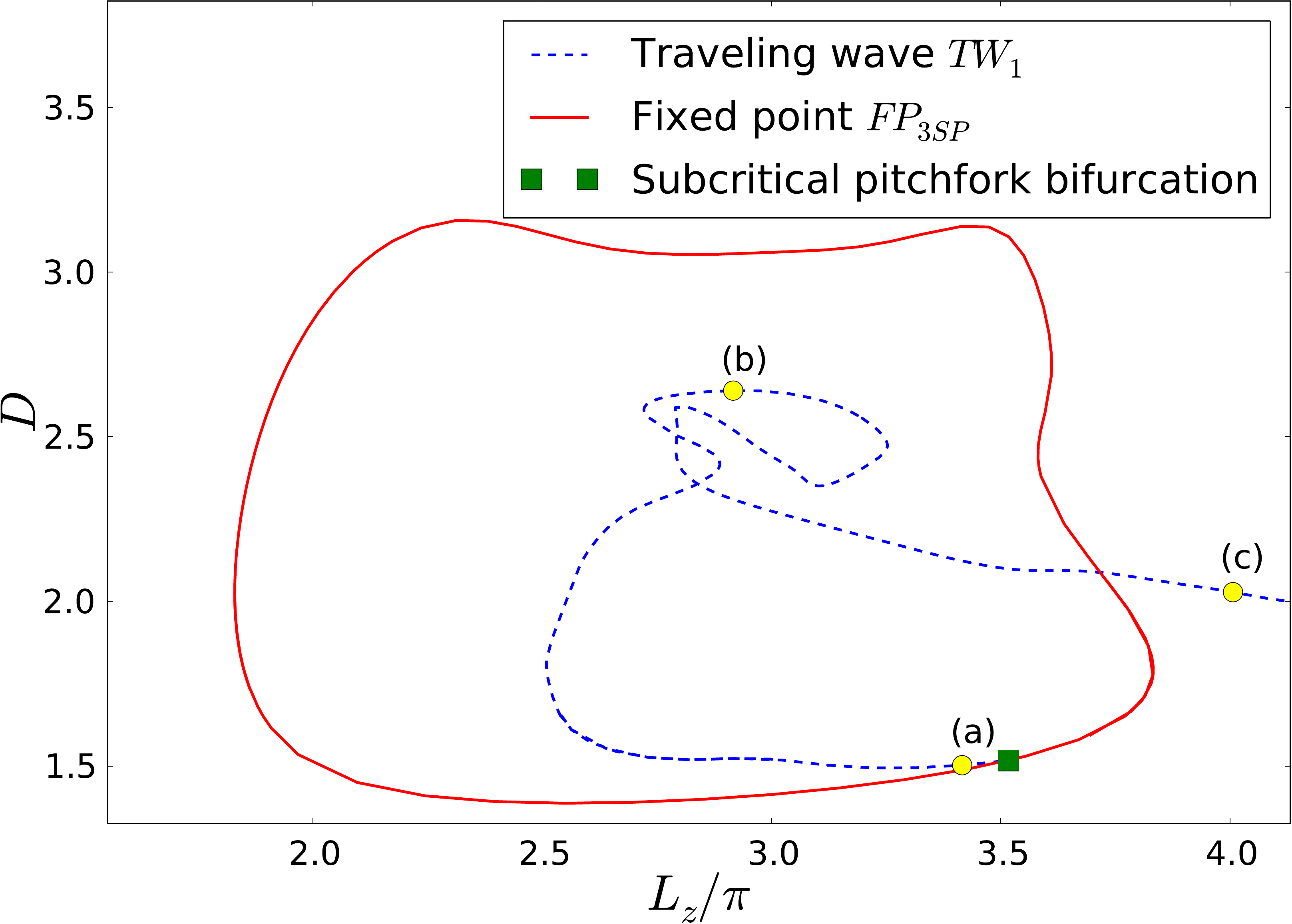}\label{ContLzFP_TW}}
\subfigure[]{\includegraphics[width=0.46\textwidth]{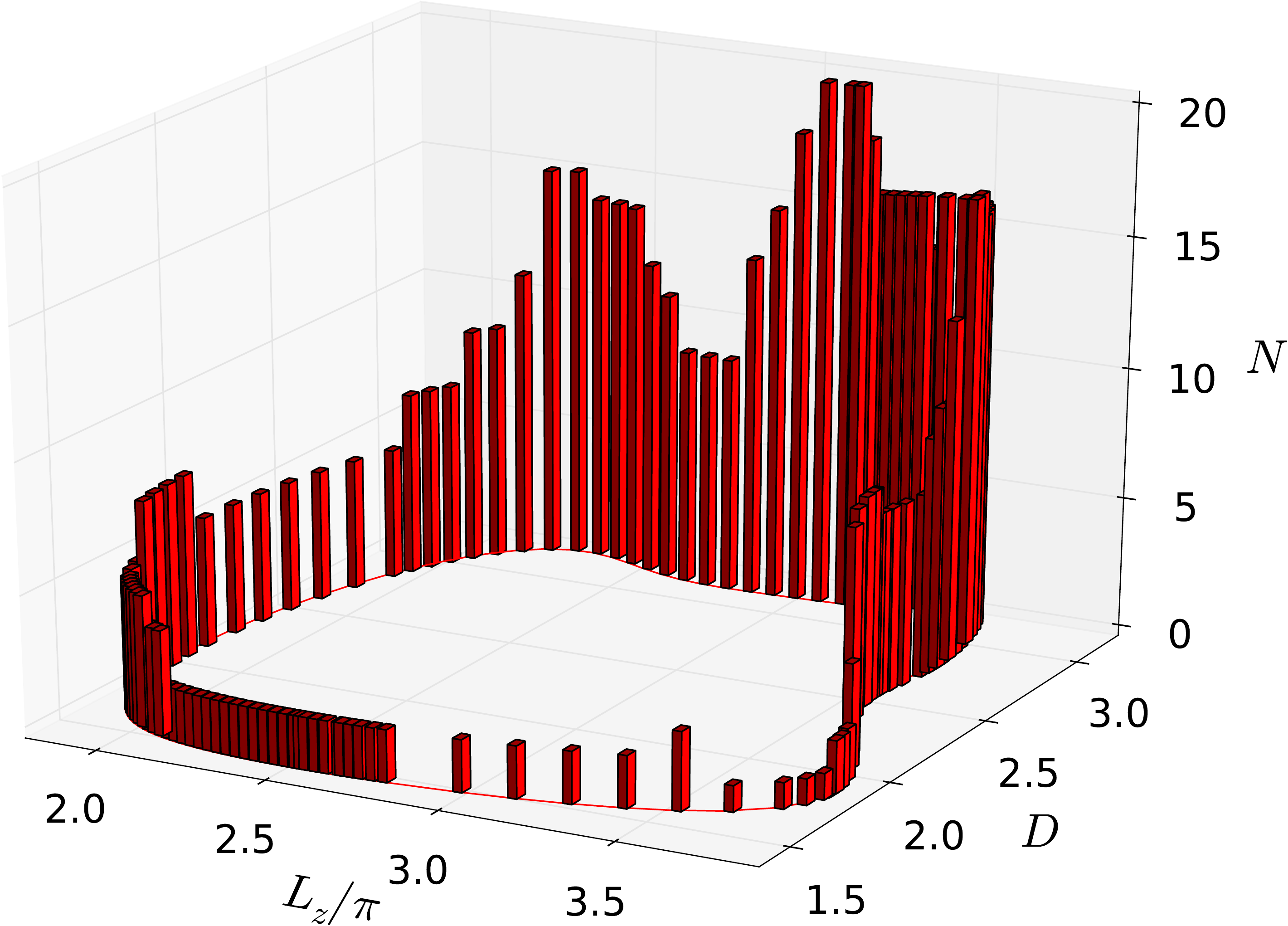}\label{Lz_FP_Stab3D}}
\caption[]{(Color online) Properties of the coherent states in wider domains: (a) dissipation $D$ as a function of spanwise period
and (b) number of unstable directions of the stationary coherent state. 
Flow fields for the states marked $(a)$ to $(c)$ are displayed in Fig.\,÷÷\ref{fig:PlotsTW}. 
The Reynolds number is  $Re=400$.}
\label{fig:ContinuationInLz}
\end{figure}

\begin{figure}
\subfigure[]{\includegraphics[width=0.4\textwidth]{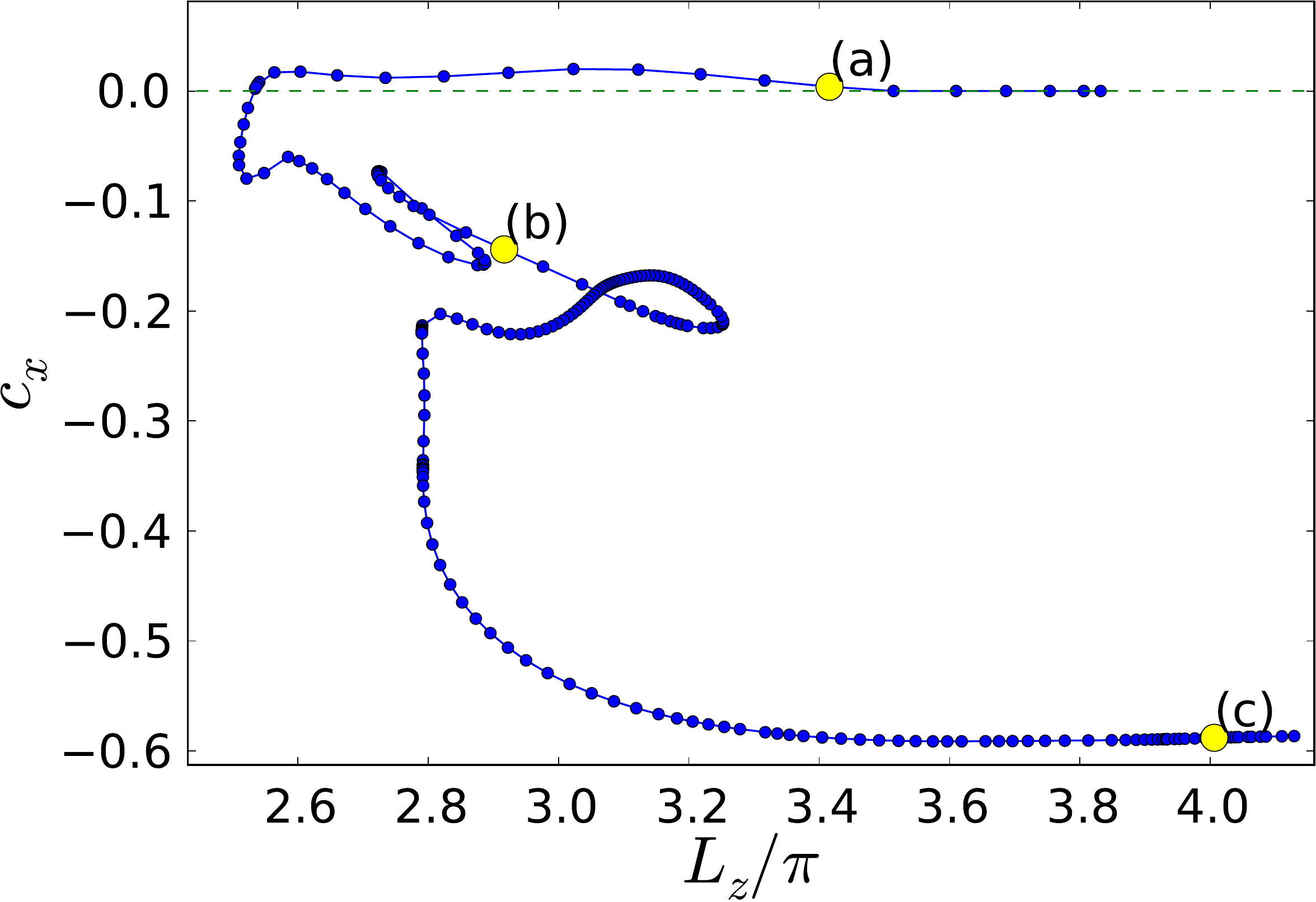}\label{VelocityTW_Lz}}
\subfigure[]{\includegraphics[width=0.4\textwidth]{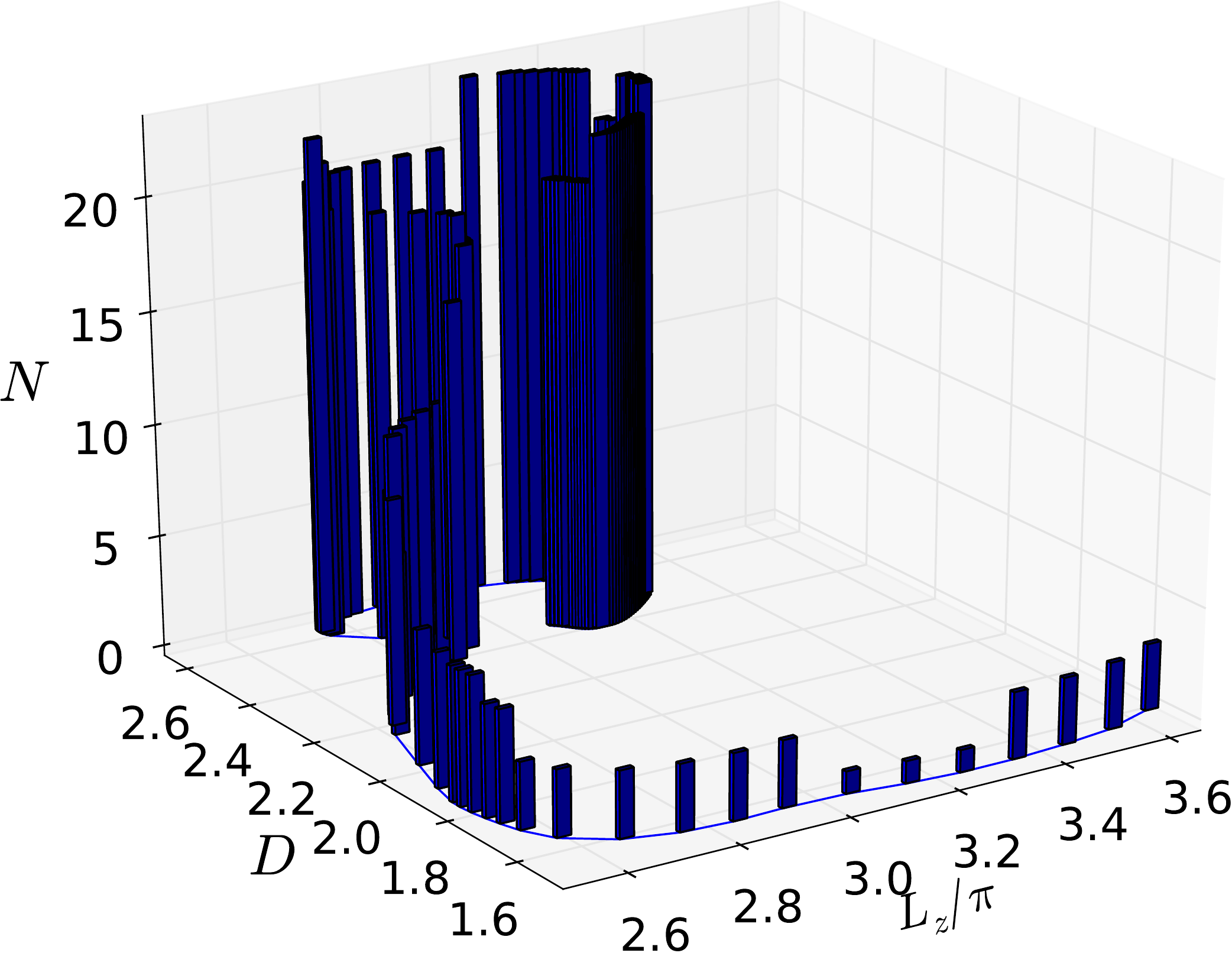}\label{Lz_TW_Stab3D}}
\caption[]{(Color online) Properties of the traveling wave shown in Fig.\,÷÷\ref{fig:ContinuationInLz}.
(a) phase speed, (b) number of positive eigenvalues for a section of the bifurcation.
For the parts not shown the number of positive eigenvalues is larger than 20. 
The states marked $(a)$ to $(c)$ are identical to the ones indicated in Fig.\,÷÷\ref{fig:ContinuationInLz};
their flow fields are shown in Fig.\,÷÷\ref{fig:PlotsTW}.}
\label{fig:TWspeedNEV}
\end{figure}

\begin{figure}
\subfigure[]{\includegraphics[width=0.45\textwidth]{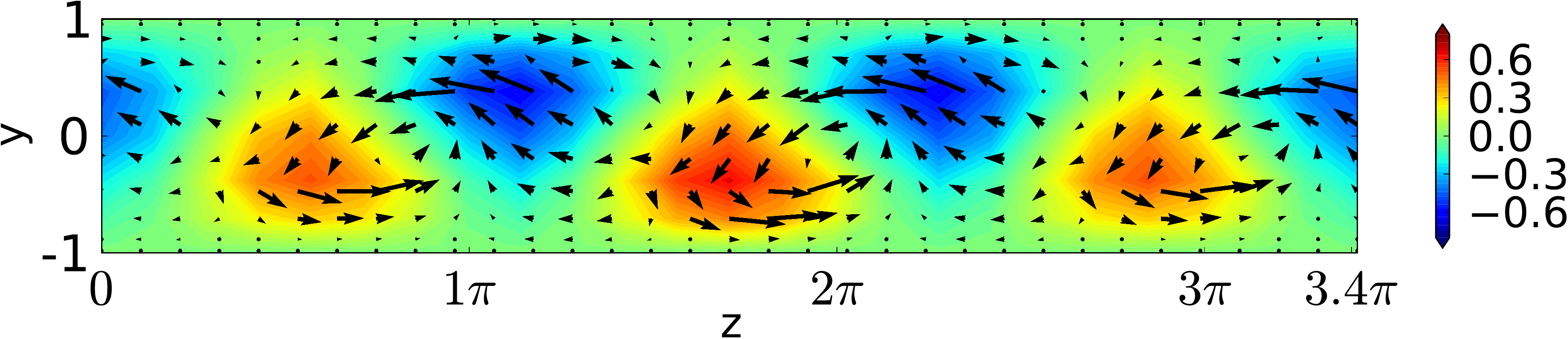}\label{PO_YZ_1}}
\subfigure[]{\includegraphics[width=0.36\textwidth]{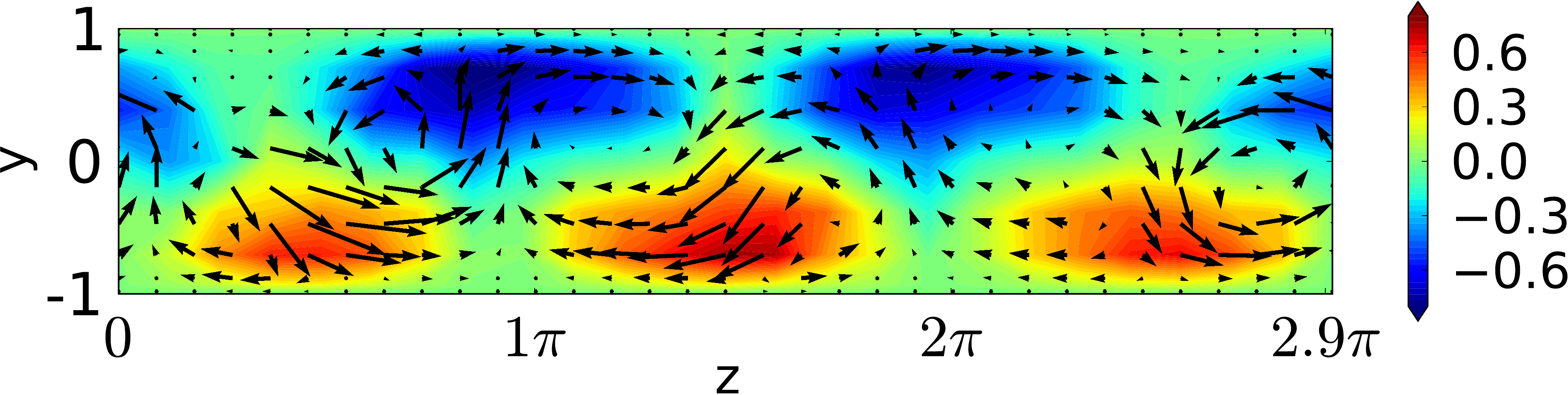}\label{PO_YZ_2}}
\subfigure[]{\includegraphics[width=0.5\textwidth]{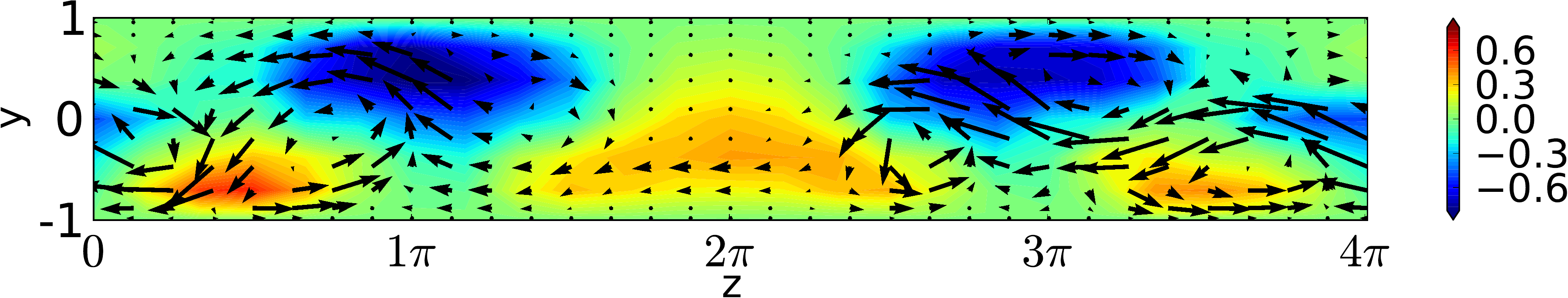}\label{PO_YZ_3}}
\caption[]{(Color online) Velocity field of the traveling wave state for the parameters marked in Fig.\,÷÷\ref{ContLzFP_TW} and \ref{VelocityTW_Lz}.
Shown are sections perpendicular to the flow direction at $x=0$, with the
in-plane velocity indicated by vectors and the perpendicular component 
indicated in color.}
\label{fig:PlotsTW}
\end{figure}


\section {Domains with three vortex pairs}
\label{sec:Domain3pi}

The state with three vortex pairs and width of $L_{z}=3\pi$ has more than one unstable
eigenvalue and hence is not a relative attractor in the laminar-turbulent boundary. 
In order to identify its role in the boundary, we employ the edge-tracking 
algorithm \cite{Toh2003,Skufca2006,Schneider2007b} to identify a nearby
edge state. In order to simplify the analysis, we impose a 
shift-reflect symmetry,
\begin{equation}
 \tau^{1/2}_{x} \sigma_{z} [u,v,w](x,y,z) = (u, v, -w)(x +0.5 L_{x}, y, -z)\,.
\end{equation} 
Within this symmetry reduced subspace, the edge-tracking algorithm quickly converges 
to a state where the up-down symmetry is broken, so that it does not correspond
to a stationary state but to a traveling wave (see Fig.\,÷÷\ref{TWPlot}). 
A linear stability analysis confirms that this
traveling wave solution indeed has only one unstable direction
in the symmetry reduced state so that it is
a relative attractor in the turbulent-laminar boundary (in the full space there are two unstable directions). 
Reflection in the midplane
gives a traveling wave that moves in the opposite direction.

In order to explore possible connections between the fixed point and the traveling wave
we first continued them in Reynolds number $Re$, see Fig.\,÷÷\ref{fig:ContinuationInRe}. 
As a measure of the amplitude of the states we show in
Fig.\,÷÷\ref{ContReFP_TW} the dissipation rate $D$ \cite{Gibson2008}, defined as
\begin{equation}
  D =  \frac{1}{V} \int_{V} (\nabla \times \textbf{u})^{2} dV,
\end{equation}
where $\textbf{u}$ is the velocity of the perturbation to the laminar state, 
vs. the Reynolds number. The fixed point can be traced until it ends in a 
saddle-node bifurcation at about $Re=164$. 
The symmetry broken traveling wave state can be traced to a saddle node
bifurcation at $Re=206$.  Tracking the number of unstable directions
one notes that the lower branch solution typically has only a single unstable
direction, but that the upper branch undergoes several bifurcations. However, 
there are no intersections between the solution branches of the fixed point and the traveling wave.

Schneider et al. \cite{Schneider2010a} described a similar spanwise localized traveling
wave in the larger box with $L_{x}=4\pi$ and $L_{z}=8\pi$, where they do find a connection 
via a symmetry-breaking bifurcation at $Re \approx 150.2$. Therefore, we study the
bifurcation diagram as a function of geometrical parameters. Fig.\,÷÷\ref{fig:ContinuationInLz}
shows the bifurcation diagram for a fixed Reynolds number $Re=400$ and 
varying box width $L_{z}$. The fixed points trace out a closed curve
consisting of an upper and a lower branch connected in saddle node bifurcations for small and
large $L_{z}$. The lower
branch is unstable with only few unstable eigenvalues while the upper branch 
undergoes a sequence of bifurcations leading to a highly
unstable state. The closed form of the solution curve allows for an easy interpretation: the state with three
vortex pairs cannot be sustained if the distance between the streaks become broader with increasing $L_{z}$,
or if the domain is too narrow and the states are squeezed together.
This has also been found in other cases \cite{Eckhardt2008}. However, as is evident in particular for
the traveling wave state, the trace of the states in the $D$-$L_{z}$-space can be quite convoluted, thereby giving
rise to a coexistence of several states for the same value of $L_{z}$ that are connected through
several saddle-node bifurcations when studied as a function of the width.

The more complicated behaviour of the traveling wave state shown in Fig.\,÷÷\ref{ContLzFP_TW} is also reflected
in other properties. The phase speed and the number of corresponding unstable eigenvalues are shown in 
Fig.\,÷÷\ref{fig:TWspeedNEV}. The phase speed changes sign from positive to negative along the
continuation curve, so that there is no fixed relation between flow speed and phase speed. It also
implies that the traveling wave becomes a stationary state without up-down symmetry for $L_{z}=2.53\pi$.
The eigenvalues shown in Fig.\,÷÷\ref{Lz_TW_Stab3D} reveal that the lower branch of the traveling wave is unstable with
one or two unstable directions only, as is the case with the stationary state. 
It has a single eigenvalue only in a small region around $L_{z}=3.1\pi$.
In contrast, the number of unstable eigenvalues for the upper branch can exceed 20 in certain parameter regions. 
The flow fields along the bifurcation curves maintain their prominent vortex and streak elements, but
change in the detailed spatial arrangements. To give 
an impression of the structural changes of the traveling wave along the continuation curve we show 
contour plots of characteristic states in Fig.\,÷÷\ref{fig:PlotsTW} for the 
three states marked by the letters $(a)$, $(b)$, and $(c)$ in Fig.\,÷÷\ref{ContLzFP_TW}
and \ref{VelocityTW_Lz}. In point $(a)$, which is close to the bifurcation point where
the stationary state and the traveling wave are connected, the traveling wave consists of three almost 
equal vortex pairs. Approaching point $(b)$ the streaks become compressed in the upper
or lower half of the domain, and tend to become a bit broader, even though the width of the 
domain is reduced to $L_{z}=2.91\pi$. They also become dynamically more active, which
causes the higher dissipation rate. Finally, in point $(c)$ for a box with $L_{z}=4.00\pi$ the 
streaks spread out some more and regions with very little vortical activity begin to appear
(for instance in the middle of Fig.\,÷÷\ref{fig:PlotsTW}c).

\begin{figure}
\includegraphics[width=0.5\textwidth]{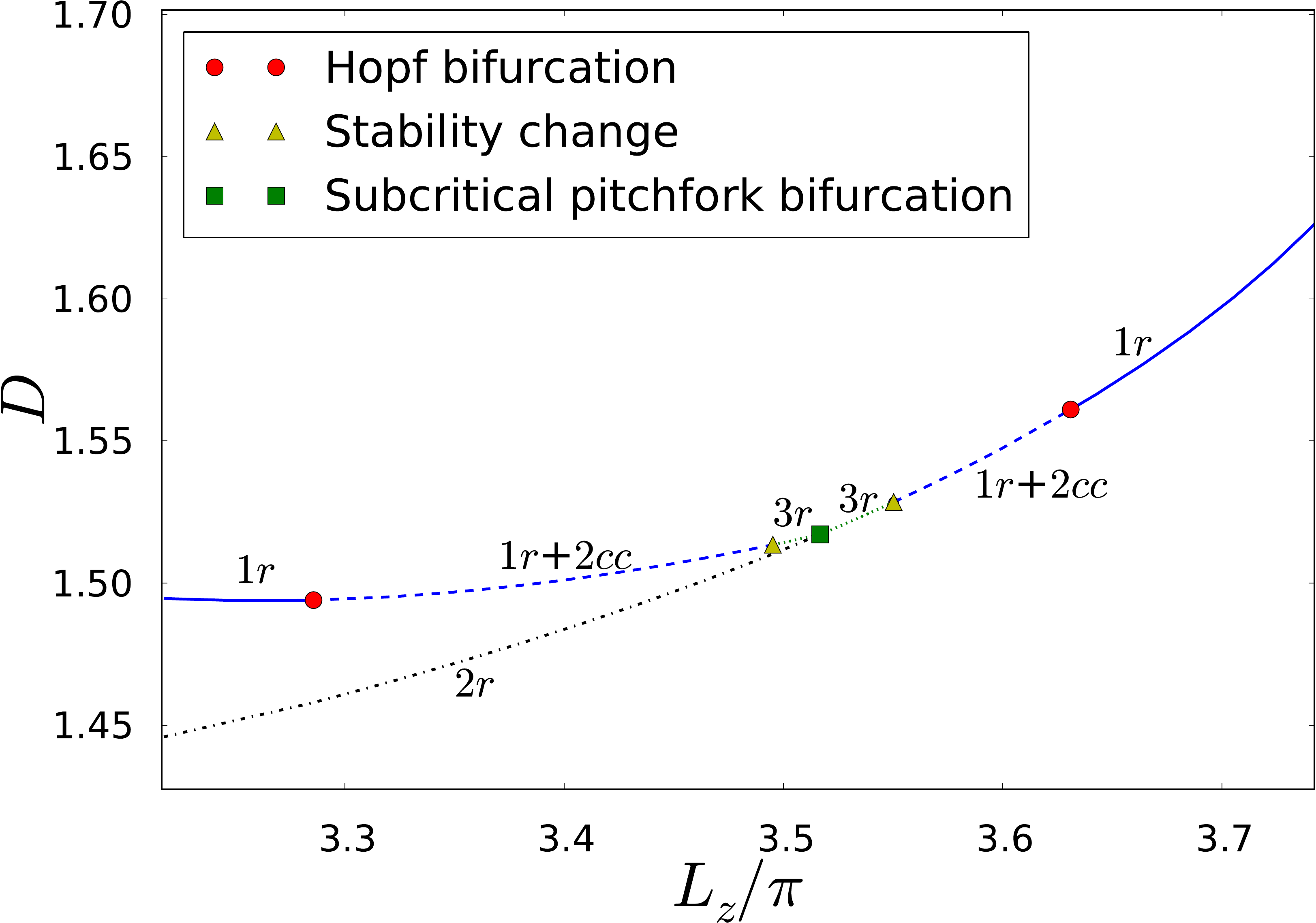}
\caption[]{(Color online) Zoom into Fig.\,÷÷\ref{ContLzFP_TW}: the traveling wave branch bifurcates from the fixed point branch
at $L_{z}=3.51\pi$ via a subcritical pitchfork bifurcation. $r$ stands for a real eigenvalue, $cc$ indicates
a complex conjugate eigenvalue. 
The dimensionality of the corresponding unstable manifold is given by numbers,
e.g. 2cc stays for a pair of complex conjugate eigenvalues.}
\label{ContLz_Bifurcations}
\end{figure}

Returning to the issue of the connection between the stationary state and the traveling wave, we note
an intersection of the fixed point branch and the traveling wave branch 
at $L_{z}=3.51\pi$, see Fig.\,÷÷\ref{ContLz_Bifurcations}. This intersection corresponds to a subcritical pitchfork
bifurcation in which the traveling wave is born. Both the emerging traveling wave branch for $L_{z}<3.51\pi$ and the fixed
point branch for $L_{z}>3.51\pi$ first undergo a stability change and then a Hopf bifurcation. In the point
where the stability changes, two positive real eigenvalues collide and emerge as a pair of complex conjugate eigenvalues. 
In the Hopf bifurcation a complex
conjugate pair of eigenvalues crosses the imaginary axis. We also note 
that both the fixed point and the
traveling wave solutions become edge states with co-dimension one stable manifolds in the corresponding 
Hopf bifurcations.
The traveling wave branch then undergoes a Hopf bifurcation at $L_{z}=3.28\pi$, while the Hopf bifurcation
of the fixed point branch occurs at $L_{z}=3.63\pi$, see Fig.\,÷÷\ref{ContLz_Bifurcations}. Both solutions
transform from an unstable state into an attractor in the laminar-turbulent boundary.

\begin{figure}
\subfigure[]{\includegraphics[width=0.42\textwidth]{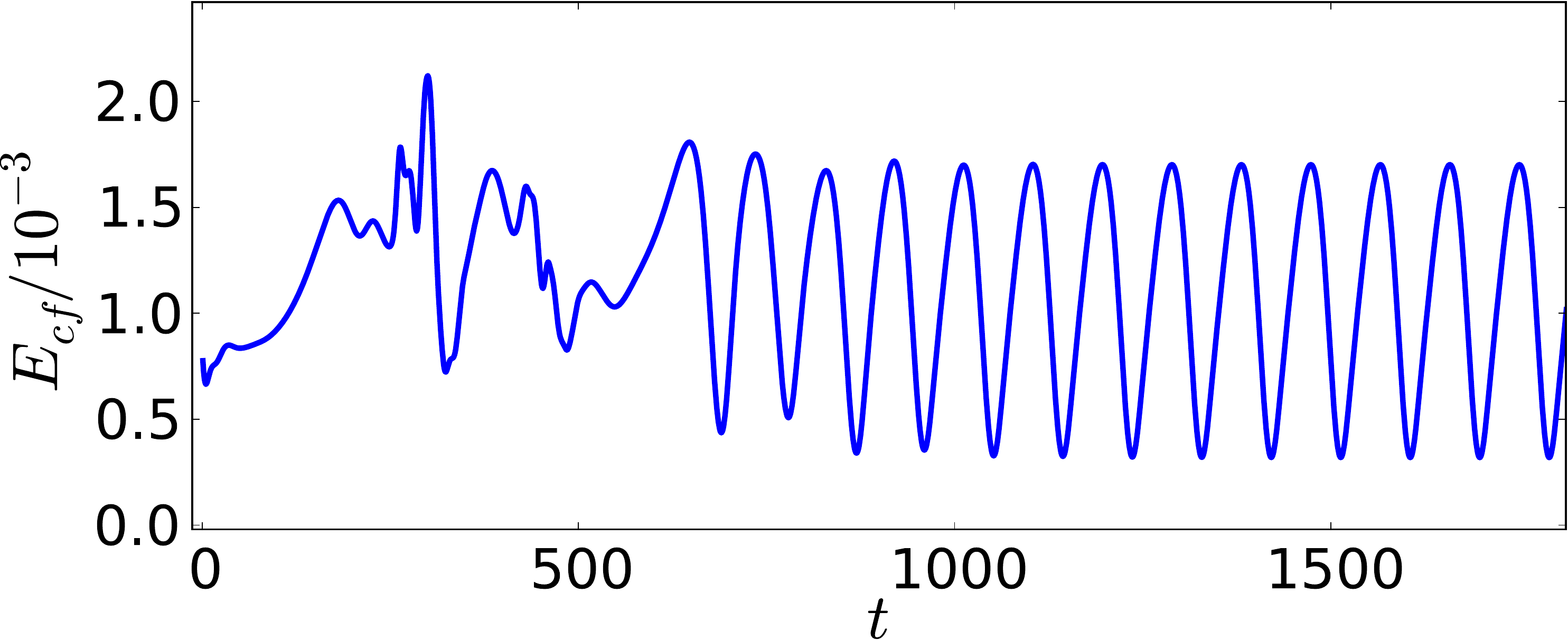}\label{PO1_EdgeTracking}}
\subfigure[]{\includegraphics[width=0.5\textwidth]{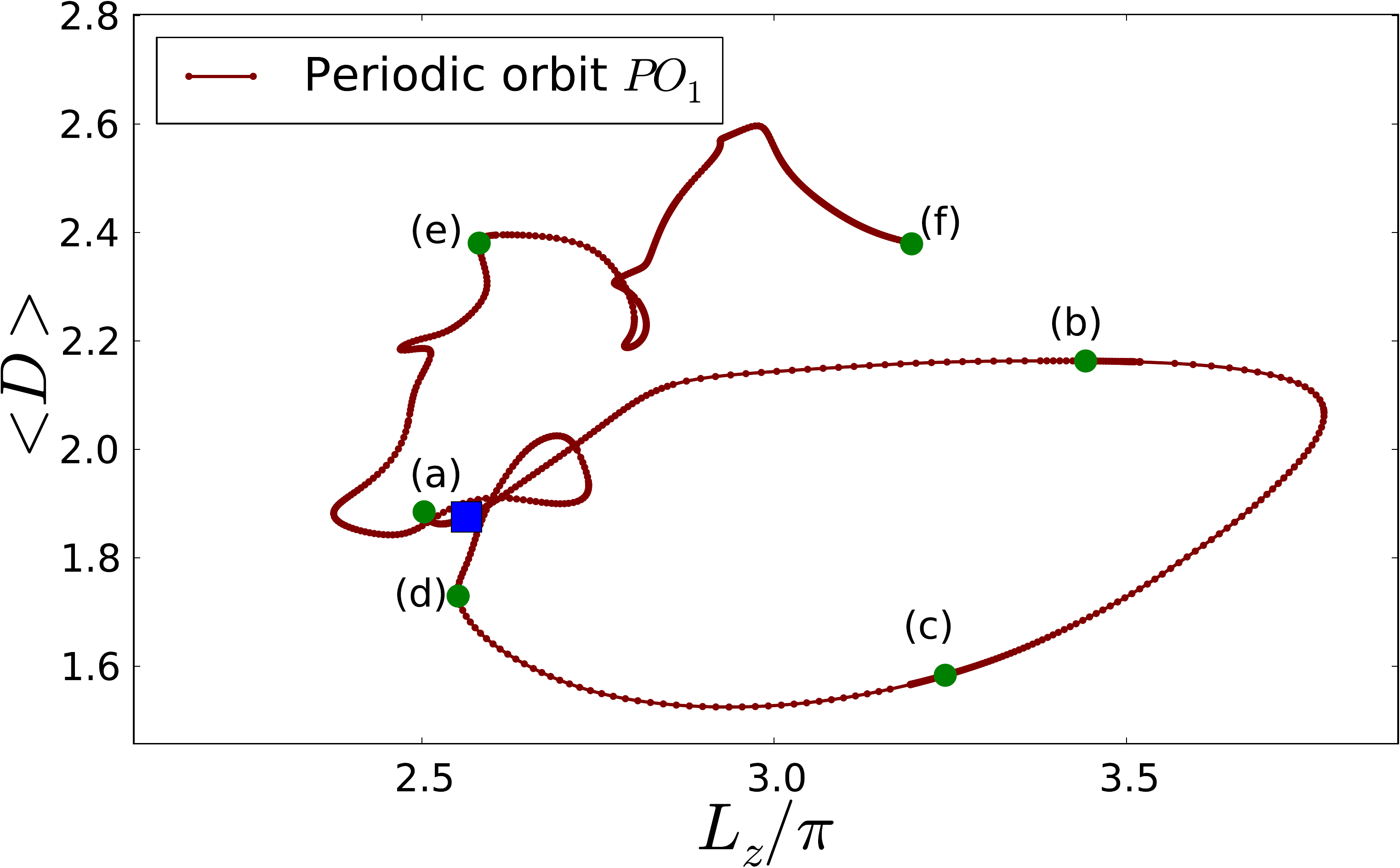}\label{PO1_LzCont}}
\caption[]{(Color online) The periodic orbit that is an edge state in this system.
(a) Cross flow energy for the edge tracking algorithm.
(b) Tracing the orbit as a function of widths, where every point corresponds to an exact numerical solution.
Labels $(a)$ to $(f)$ mark solutions which are plotted in Fig.\,÷÷\ref{fig:PO_Lz_3Dplots}, the square marks the first
solution that was found by the edge tracking algorithm.}
\end{figure}

The linear stability analysis shows that in the box with width $L_{z}=2.56\pi$ neither the traveling wave nor the
fixed point are edge states. We, therefore, used the edge tracking algorithm to pin down a nearby edge state.
Using the cross flow energy 
\begin{equation}
  \label{DefEcf}
  E_{cf} = \frac{1}{L_{x}L{z}} \int (v^{2}+w^{2}) dV
\end{equation}
as a measure of the amplitude of the state we find that the edge tracking algorithm converges to a periodic signal,
see Fig.\,÷÷\ref{PO1_EdgeTracking}. A subsequent Newton search then converged to a periodic orbit with 
period $T_{full}$ = 184.72. A visualization of the flow field shows that it is shifted downstream by half a box length after
half a period. In order to find possible connections to other solutions, 
this periodic orbit was continued in the box width $L_{z}$. 
The results are shown in Fig.\,÷÷\ref{PO1_LzCont} where every point corresponds to an exact numerical solution
found with the Newton-hookstep algorithm. In order to trace the continuation of this periodic orbit in parameter space,
we calculate the dissipation rate averaged over one period $\langle D \rangle$,
\begin{equation}
 \langle D \rangle =  \frac{1}{T_{p}} \int_{0}^{T_{p}} D(t) dt.
\end{equation}
The square in Fig.\,÷÷\ref{PO1_LzCont} marks
the initial box width of $L_{z}=2.56\pi$. It turned out to be difficult to trace this state to narrower boxes because
its period grows very quickly. A continuation towards increasing
box widths is far simpler since the period decreases and eventually stabilizes around $T=10$. The continuation
curve shows
a quite complicated behavior, with many loops and several parameter ranges with multiple
upper and lower branches.

\begin{figure}
\centering
\subfigure[]{\includegraphics[width=0.1877\textwidth]{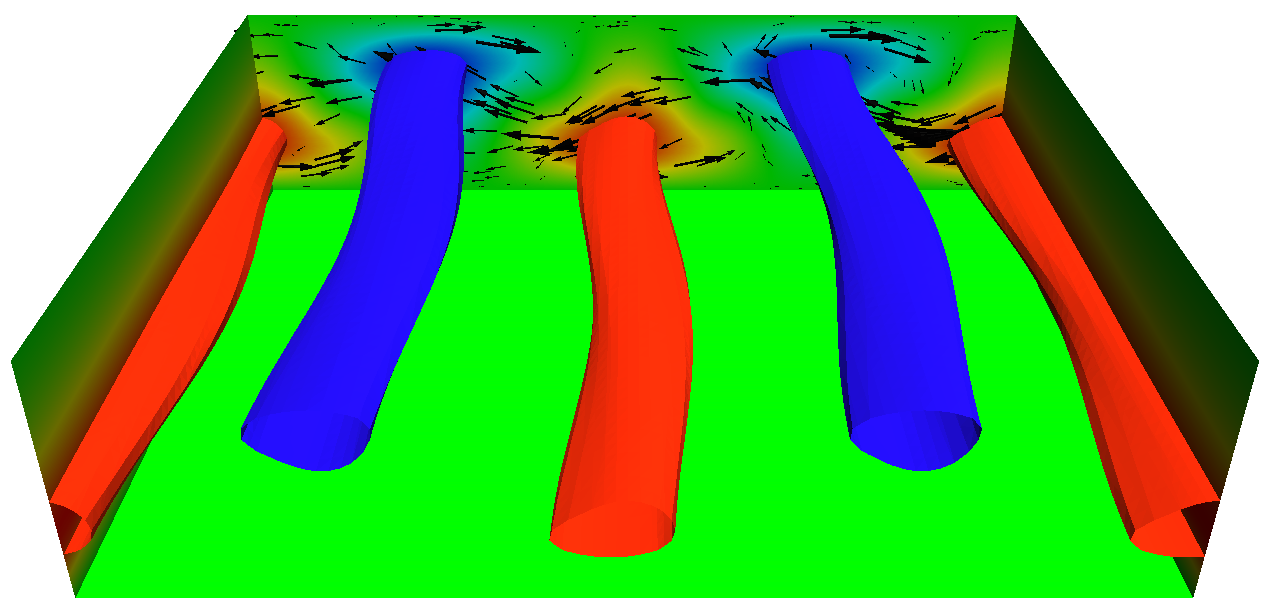}\label{PO_3D_1}}
\hfill
\subfigure[]{\includegraphics[width=0.2488\textwidth]{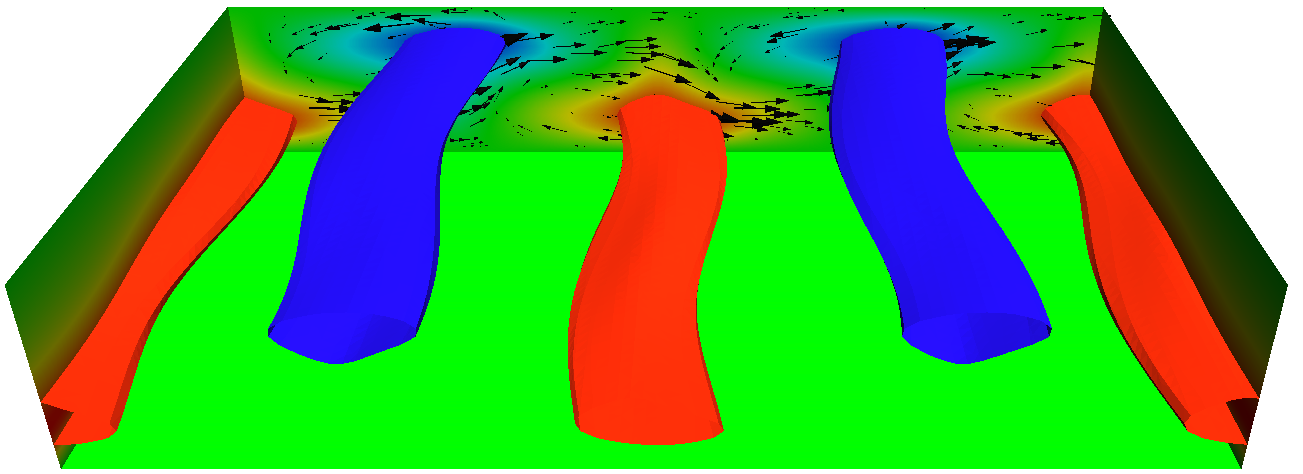}\label{PO_3D_2}}
\hfill
\subfigure[]{\includegraphics[width=0.23658\textwidth]{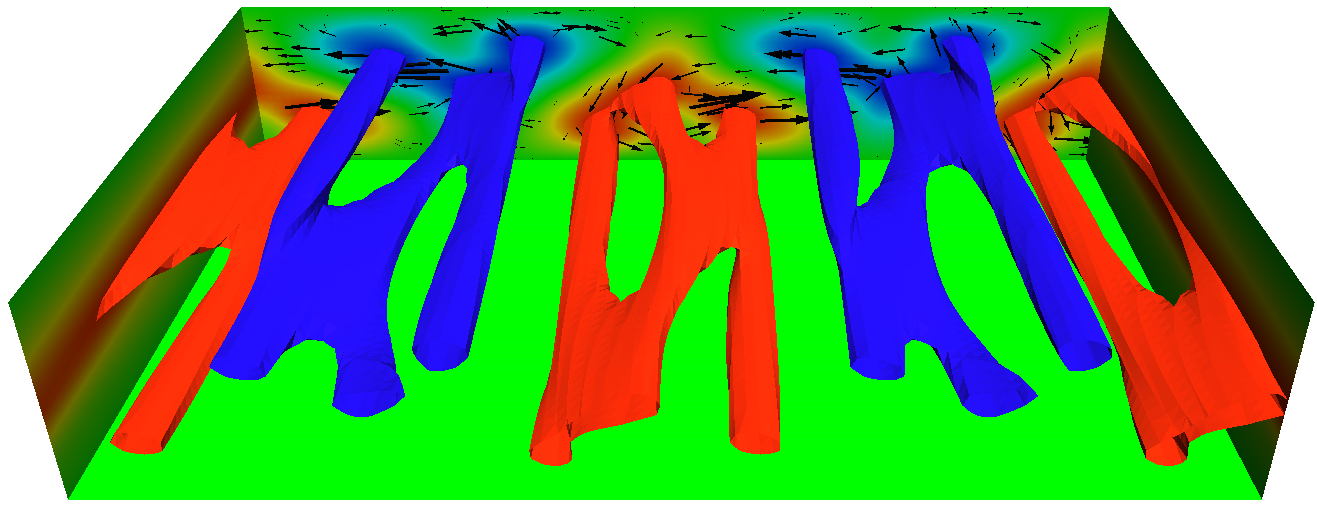}\label{PO_3D_3}}
\hfill
\subfigure[]{\includegraphics[width=0.19992\textwidth]{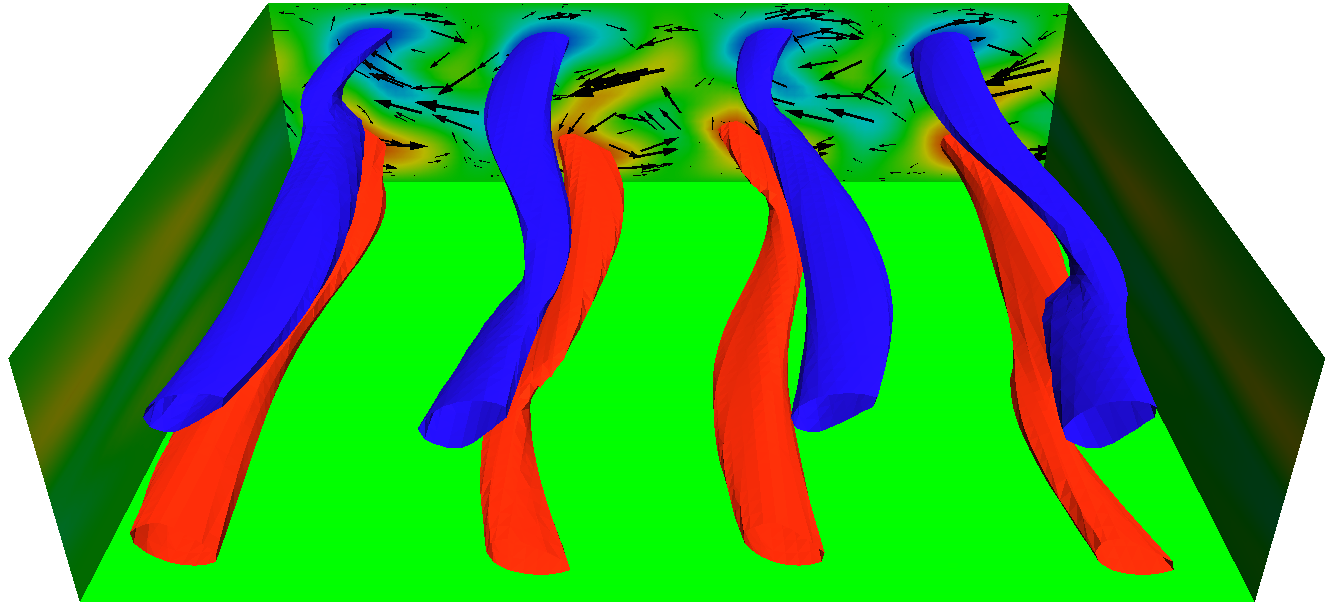}\label{PO_3D_4}}
\hfill
\subfigure[]{\includegraphics[width=0.20079\textwidth]{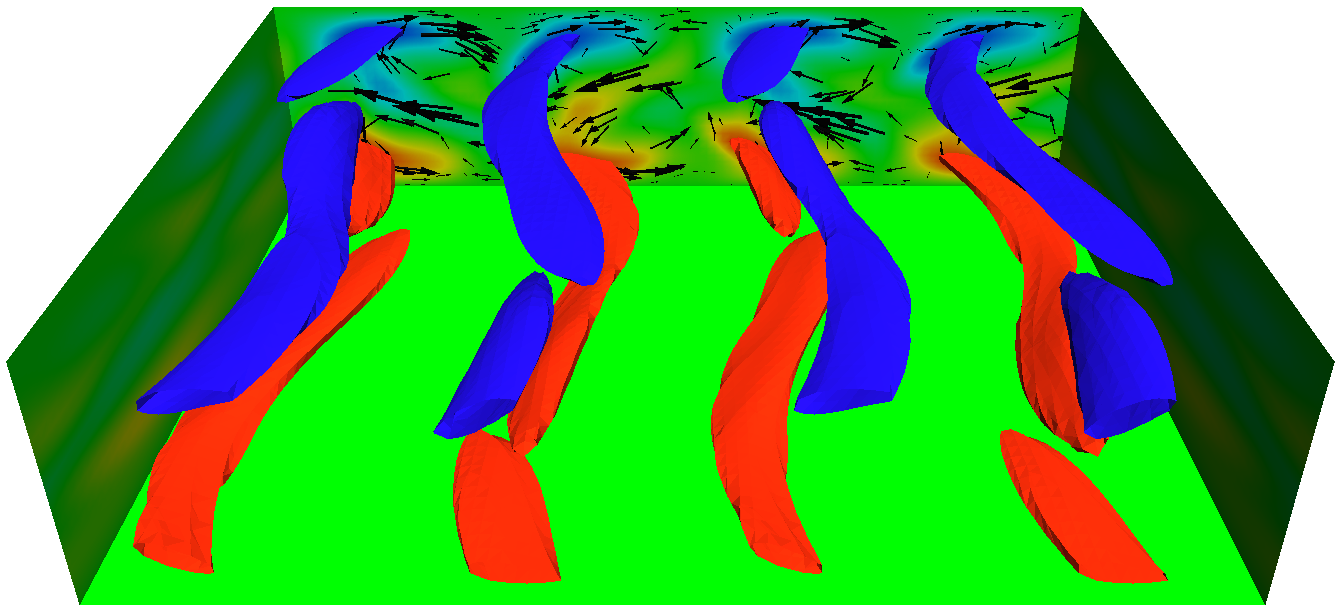}\label{PO_3D_5}}
\hfill
\subfigure[]{\includegraphics[width=0.23571\textwidth]{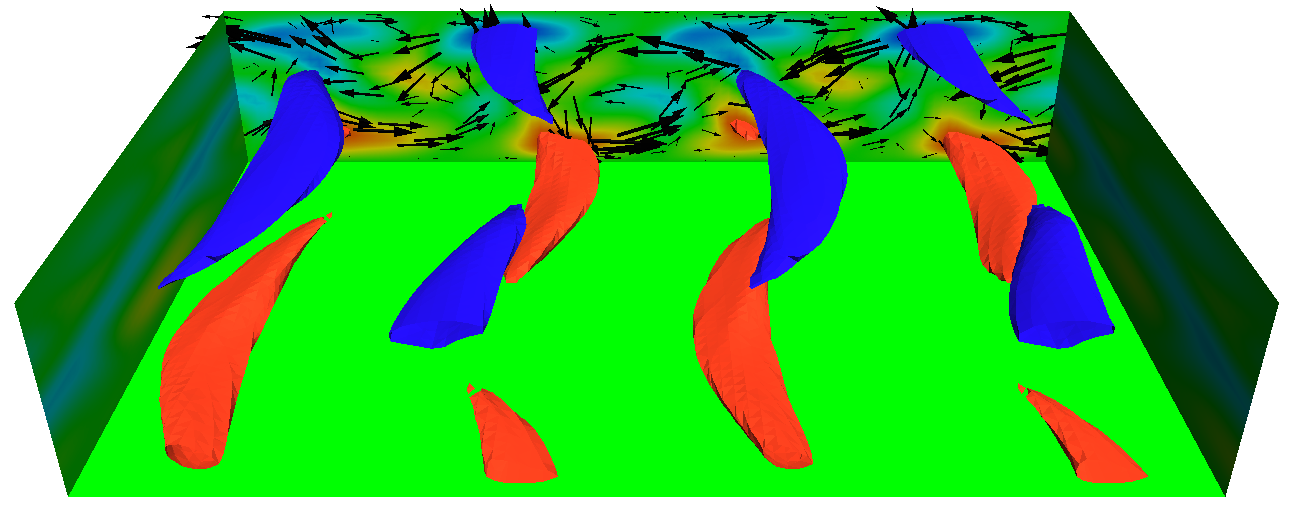}\label{PO_3D_6}}
\caption[]{(Color online) Snapshots of the periodic orbit $PO_{1}$ at parameter points marked along the continuation curve in Fig.\,÷÷\ref{PO1_LzCont}.
All states are shown 
at the minimal energy of the periodic orbit. The prominent red (lower) and blue (upper) regions are iso-conturs of 
the downstream velocity at the values $u_{high, low}$ and highlight streaks. (a) $u_{high,low}$= $\pm$ 0.55, $L_{z}$ = 2.50$\pi$;
(b) $u_{high,low}$= $\pm$ 0.70, $L_{z}$ = 3.44$\pi$; (c) $u_{high,low}$= $\pm$ 0.45, $L_{z}$ = 3.24$\pi$; (d) $u_{high,low}$= $\pm$ 0.35, $L_{z}$ = 2.55$\pi$;
(e) $u_{high,low}$= $\pm$ 0.45, $L_{z}$ = 2.58$\pi$; (f) $u_{high,low}$= $\pm$ 0.45, $L_{z}$ = 3.20$\pi$.
The colors and vectors
on the side of the boxes indicate the local in-plane velocity (arrows) and the transverse velocity component (colors).
}
\label{fig:PO_Lz_3Dplots}
\end{figure}

The complicated trace in the state spaces is accompanied by noticeable changes in the periodic orbit. 
Snapshots of characteristic states that are marked in Fig.\,÷÷\ref{PO1_LzCont}
are shown in Fig.\,÷÷\ref{fig:PO_Lz_3Dplots}. 
The snapshots are taken at a point along the period where the energy of the states is minimal. 
While the periodic orbits in Fig.\,÷÷\ref{PO_3D_1} and Fig.\,÷÷\ref{PO_3D_2} are quite similar to the initial flow field (only the 
widths of the streaks has increased in the wider box), more structural changes occur if we follow the continuation 
curve to the solution (c) (Fig.\,÷÷\ref{PO_3D_3}) in the lower branch. Here, the streaks begin to split into two regions that
are connected through narrow regions only. For the parameters of Fig.\,÷÷\ref{PO_3D_4} these connections are almost
lost, and the flow is dominated by four streak pairs.  The transformation goes even further: 
the streaks also start to divide in the streamwise direction, producing fully localized structures as can be seen
in Figs.÷\ref{PO_3D_5} and \ref{PO_3D_6}. 

\begin{figure}
\includegraphics[width=0.5\textwidth]{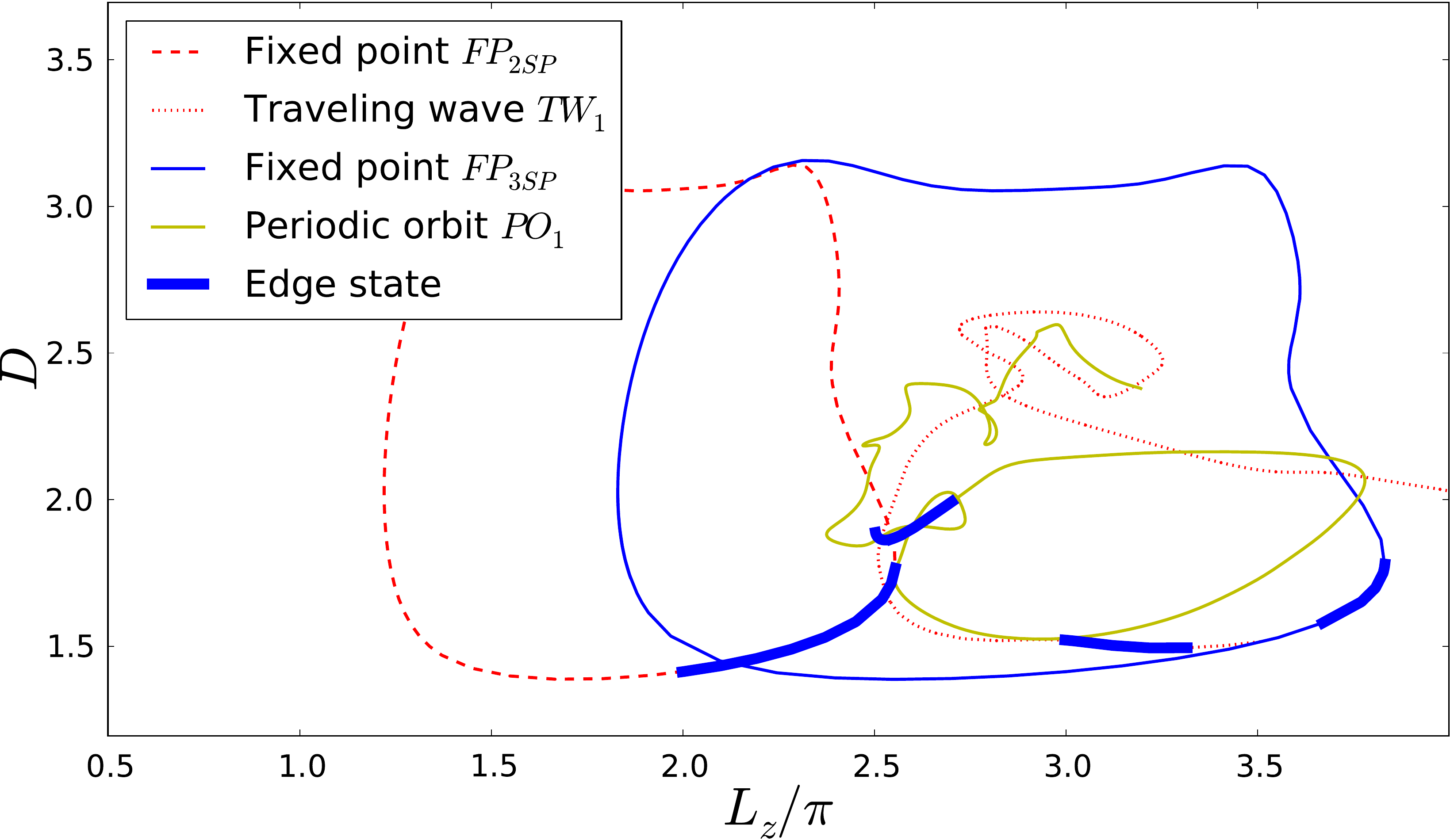}
\caption[]{(Color online) Summary of the dynamically relevant solutions in the region around $L_{z}=3\pi$. Shown is the
dissipation rate $D$, and in cases of periodic orbits the averaged dissipation rate $\langle D\rangle $ vs length $L_{z}$. States
that have only one unstable direction in the full phase space and hence are edge states, are marked by heavy lines. 
}
\label{Lz_EdgeStatesOverview}
\end{figure}

A summary of all bifurcations and continuations of stationary states, traveling waves and periodic solutions is given in 
Fig.\,÷÷\ref{Lz_EdgeStatesOverview}. 
The thick lines mark the regions where the corresponding states are edge states. 
One notes that the fixed point with three streak pairs $FP_{3SP}$ is an edge state in the range 
$3.66\pi<L_{z}<3.82\pi$. It looses its stability shortly after the saddle-node
bifurcation of the branch in a collapse of two real eigenvalues which produce a pair of complex conjugate eigenvalues. 
The periodic orbit $PO_{1}$ is an edge state for $2.51\pi<L_{z}< 2.67\pi$. 
For increasing $L_{z}$ it undergoes a Hopf bifurcation and 
a pair of complex conjugate eigenvalues emerge. The diagram shows that the fixed point with two streak 
pairs $FP_{2SP}$ is an edge state in the same region as the periodic orbit. 
%



\section{Localization in Wider boxes}
\label{sec:WideBoxes}
The instabilities that appear in domains with three vortex pairs give rise to new states that show a modulation in the
spanwise direction. To see how these modulations evolve in even wider domains, we 
here use edge tracking to study the long wavelength instabilities in
boxes with $L_{z}=5\pi$ and $L_{z}=7\pi$. The initial fields consist of the periodically 
repeated small domain state, perturbed in the direction of the second unstable eigenfunction.

\begin{figure}
\subfigure[]{\includegraphics[width=0.25\textwidth]{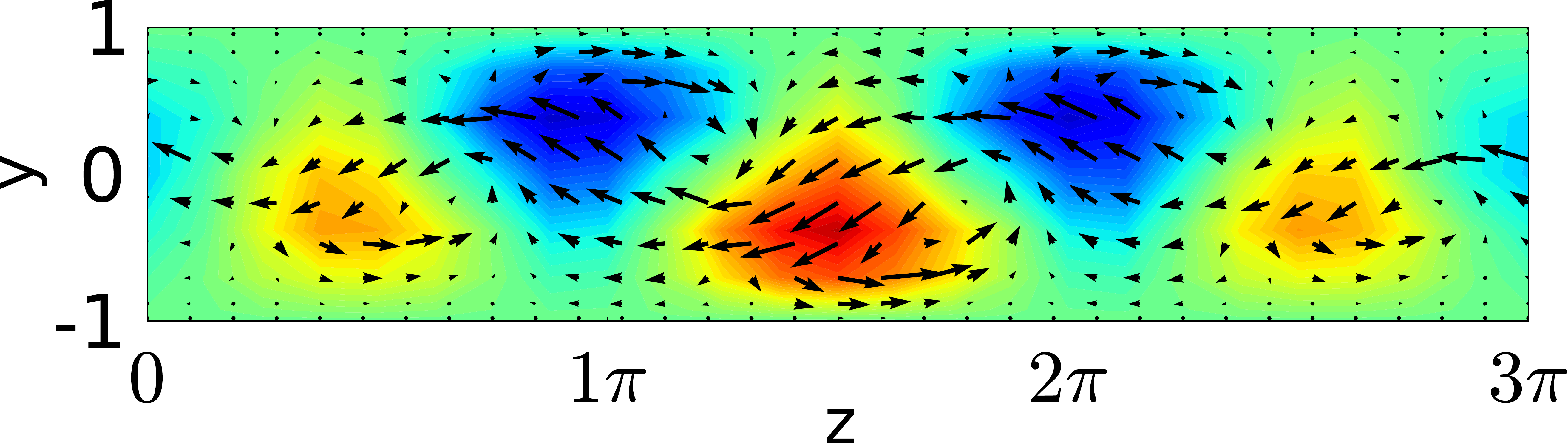}\label{WideBoxes3pi}}
\subfigure[]{\includegraphics[width=0.38\textwidth]{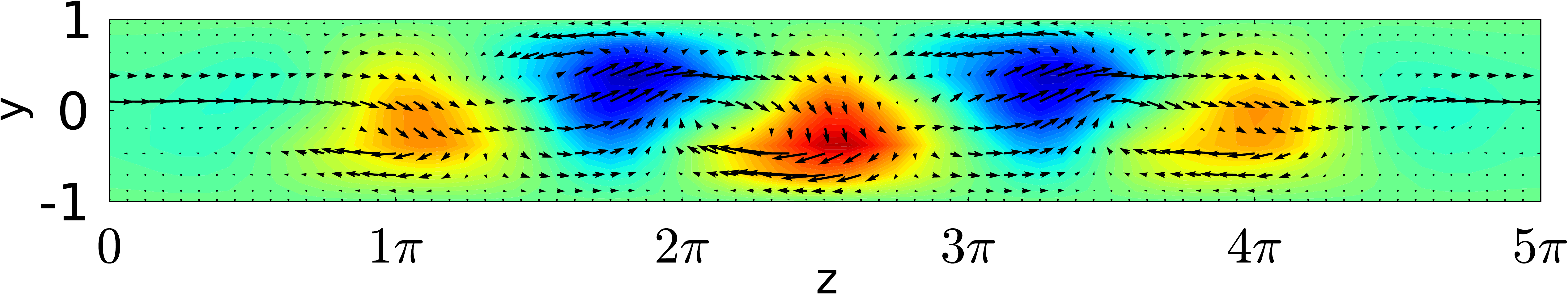}\label{WideBoxes5pi}}
\subfigure[]{\includegraphics[width=0.5\textwidth]{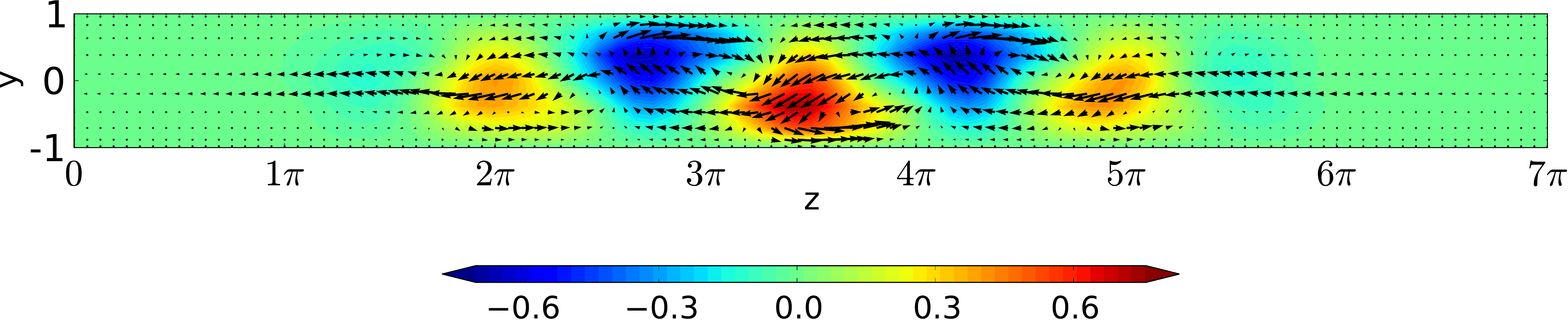}\label{WideBoxes7pi}}
\caption[]{(Color online) Cross sections of the travelling wave in wider and wider boxes. Shown are sections
perpendicular to the flow direction at $x=0$. 
As before, in-plane velocity components are indicated by vectors, 
out of plane components in color. The mean velocity profile has been subtracted.}
\end{figure}

In the case of $L_{z}=5\pi$ the edge tracking algorithm approached a stationary solution and remained in its
neigbhourhood for a while before moving away. A Newton-hookstep algorithm starting from this transiently visited
state converged to a traveling wave that was localized in the spanwise direction, see Fig.\,÷÷\ref{WideBoxes5pi}. 
This wave has two unstable directions in the symmetry constrained subspace, so that eventually the edge trajectory
is driven away from it. In the even wider box with $L_{z}=7\pi$ edge tracking converged. 
A subsequent Newton search also converged to the localized traveling wave shown in Fig.\,÷÷\ref{WideBoxes7pi}. 
This solution has only one unstable direction and thus is an edge state. 

\begin{figure}
\includegraphics[width=0.5\textwidth]{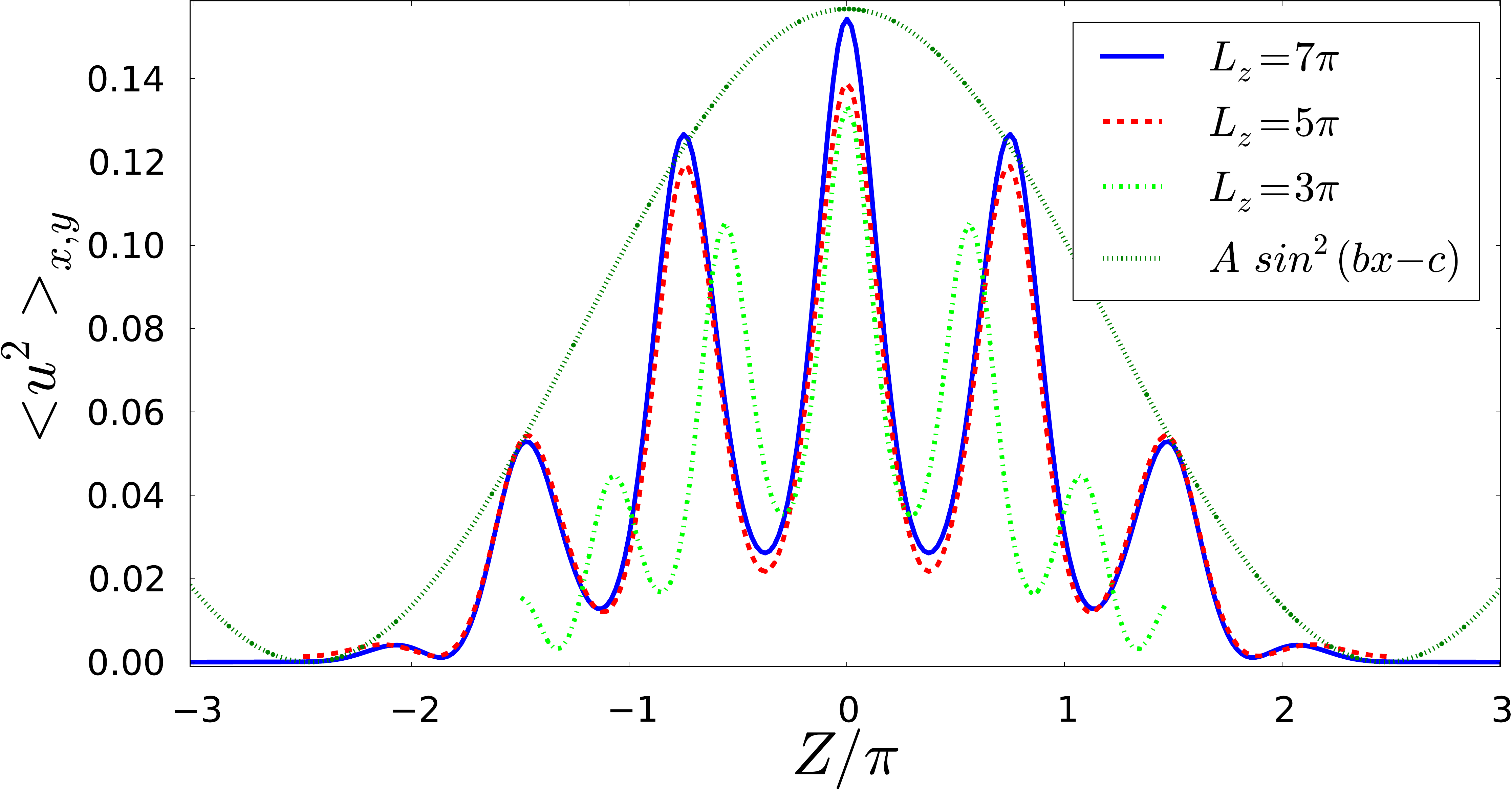}
\caption[]{(Color online) Convergence of the states as the domains become wider.
Shown is the energy content of the field $\langle u^{2}\rangle $  (without the laminar profile) vs spanwise position $z$,
together with an envelope represented by $\sin^{2}$.
}
\label{Profiles_3_5_7}
\end{figure}

The localization process of the traveling wave in domains of increasing width is summarized in Fig.\,÷÷\ref{Profiles_3_5_7}.
In the domain with $L_{z}=3\pi$ one notes that the central region of the new state is enhanced, 
for $L_{z}=5\pi$ the localization is apparent
and for $L_{z}=7\pi$ it is essentially complete. Note that the width of the state, as measured, say, by the distance between
the outermost streaks, increases from (a) to (b) and then decreases slightly again in going to (c). The contraction 
as the states
become more  localized has been seen in other situations as well \cite{Schneider2010} and seems to be a general
feature. This then suggests that the unusual stability properties of the $L_{z}=5\pi$-state are a consequence of the
geometrical constraints and atypical.

The strong localization of the states in domains of increasing width is further demonstrated in
Fig.\,÷÷\ref{Profiles_3_5_7}, where the envelope of the downstream energy content is fitted to 
a function of the form $f(z)=A \sin^{2}(bz-c)$. The particular functional form is not so relevant, 
but it highlights the strong drop off in the spanwise direction. Also, the alignment between the intensities in
domains of width $L_{z}=5\pi$ and $7\pi$ shows that the states are already close to their asymptotic form
for infinitely wide domains.

\section{Concluding remarks}
\label{sec:Conclusion}
The stability analysis of exact coherent structures in wide domains presented here provides an 
approach to the determination of localized structures via the use of the edge tracking algorithm. 
It shows that extended states that are multiple copies of the
exact coherent structures from small domains develop long-wavelength instabilities that break
the translational symmetry. As a consequence, new states that are modulated in the spanwise direction
develop. Increasing the width of the domains further the
modulations are amplified until the states are fully localized. The details of the bifurcations turn out
to be rather complex, as each state involved in the analysis has a rich bifurcation diagram when
analyzed as a function of domain width. While the bifurcation diagram in Reynolds number
that is shown in Fig.\,÷÷\ref{ContReFP_TW} is rather simple, one has to keep in mind that in the
combined space of all parameters, which include in addition the Reynolds number also the width
and length of the domain, many more combinations and bifurcations should be expected. 

While the details of the bifurcation diagram will depend on the particular system and parameters
being analysed, the overall features should be  fairly robust. In particular the development of 
long-wavelength instabilities as a mechanism to generate localized structures out of periodically extended
ones can be expected to occur in other systems as well. We therefore expect that
similar bifurcation patterns can be found in other shear flows where localized structures appear, 
such as the boundary layer cases studied in \cite{Cherubini2011,Biau2012,Duguet2012}
or the localized edge states in the asymptotic boundary layer, \cite{Khapko2013, Kreilos2013}.

Finally, we would like to point out that the long-wavelength stability analysis can also be performed 
in the downstream direction. 
One should then find transitions to solutions that are periodically extended in the spanwise
direction but localized downstream \cite{Marinc2010,Avila2013} or fully localized in both directions 
\cite{Schneider2010a,Mellibovsky2009}.

Acknowledgement: 
This work was supported in part by the Deutsche Forschungsgemeinschaft within
Forschergruppe 1182. 

%

\end{document}